\documentstyle[12pt,epsf,epsfig,wrapfig]{article}
\newcommand{ \bm}[1]{\mbox{\boldmath{$#1$}}}

\begin{document}
 \begin{center}
 {\bf POLARIZATION PHENOMENA IN FRAGMENTATION OF DEUTERONS TO PIONS
 AND NON-NUCLEON DEGREES OF FREEDOM IN THE DEUTERON
}

 \vskip 5mm
 A.Yu.~Illarionov$^{\dag}$, A.G.~Litvinenko and G.I.~Lykasov
 \vskip 5mm
 {\small
 {\it
 Joint Institute for Nuclear Research,
 141980 Dubna, Moscow Region, Russia
 }
 \\
 $\dag$ {\it
 E-mail: Alexei.Illarionov@jinr.ru
 }}
 \end{center}
 \vskip 5mm
 
 \begin{abstract}
  The fragmentation of deuterons into pions emitted forward in the
  kinematic region forbidden for free nucleon-nucleon collisions
  is analyzed. It is shown that the inclusion of the non-nucleon 
  degrees of freedom in a deuteron allows to describe the experimental 
  data about inclusive pion spectrum rather satisfactory and improves the 
  description of data concerning the deuteron analyzing power $T_{20}$. 
  The experimental data show the positive sign and very small values, 
  less than $0.2$, of $T_{20}$ what can't be reproduced by the 
  calculations ignoring these degrees of freedom.
 
 \end{abstract}
 
 \vskip 10mm
 
 Let us analyze the inclusive process  $D p\rightarrow\pi X$ for the
 polarized deuteron and pions emitted to forward at initial energies 
 of order few GeV.
 If the deuteron is tensor analyzed and has $p_D^{ZZ}$ component,
 then the inclusive spectrum of this reaction can be written in the form:
 \begin{equation}
 \rho_{pD}^\pi\left(p_D^{ZZ}\right)=
 \rho_{pD}^\pi \Bigl[1 + {\rm A}_{ZZ} \cdot p_D^{ZZ} \Bigl]~,
 \label{def:AZZ}
 \end{equation}
 where
 $\rho_{pD}^\pi \equiv \varepsilon_\pi \cdot {d\sigma_{pD}^\pi / d^3p_\pi}$
 is the inclusive spectrum for the case of unpolarized
 deuterons and ${\rm A}_{ZZ} = \sqrt{2}T_{20}$ is the tensor analyzing
 power. They can be  written in a fully covariant manner within
 the Bethe-Salpeter formalism \cite{ill00}:
 \begin{eqnarray}
 &&\rho_{pD}^\pi =
 {1\over(2\pi)^3} \int {\sqrt{\lambda(p,n)}\over\sqrt{\lambda(p,D)}}~\left[
 \rho_{pN}^\pi\cdot\Phi^{(u)}(|\bm q|)\right]~{d^3q\over E_{\bm q}}~;
 \label{un.cs} \\
 &&\rho_{pD}^\pi\cdot{\rm A}_{ZZ} = -
 {1\over(2\pi)^3} \int {\sqrt{\lambda(p,n)}\over\sqrt{\lambda(p,D)}}~\left[
 \rho_{pN}^\pi\cdot\Phi^{(t)}(|\bm q|)\right]
 \left({3 \cos^2\theta_{\bm q} - 1 \over 2}\right)
~{d^3q\over E_{\bm q}}
 \label{calc:AZZ}
 \end{eqnarray}
 where $\lambda(p_1, p_2) \equiv (p_1p_2)^2 - m_1^2 m_2^2 =
 \lambda(s_{12}, m_1^2, m_2^2) / 4$ is the flux factor; $p, n$ are
 the four-momenta of the proton-target and intra-deuteron nucleon,
 respectively;
 $\rho_{pN}^\pi$ is the relativistic invariant inclusive spectrum of
 pions produced by interacting the intra-deuteron nucleon with the
 proton-target.
 The functions $\Phi^{(u)}(|\bm q|)$ and  $\Phi^{(t)}(|\bm q|)$ depends on
 the relative momentum $q = (n - p')/2$ and contains full information
 about the structure of deuteron with one on-shell nucleon.

 According to \cite{lyk93}, the deuteron structure can be described by
 assuming the possible existence of non-nucleon or quark degrees of freedom.
 On the other hand, the shape of a high momentum tail of the nucleon
 distribution in the deuteron can be found from the true Regge asymptotic
 at $x\rightarrow 2$ \cite{efr88}, and the corresponding parameters can be
 fitted from a good description of the inclusive proton spectrum in
 deuteron fragmentation $D p\to p X$ \cite{lyk93,efr88}. 
 According to \cite{lyk93,efr88}, one can  write the following form for
 $\Phi^{(u)}(|\bm q|)$:
 \begin{equation}
 \Phi^{(u)}(|\bm q|) = {E_{\bm k}/E_{\bm q} \over 2(1 - {\rm x})}
 \widetilde\Phi^{(u)}(|\bm k|)~.
 \label{Phi:6q-(q->k)}
 \end{equation}
where
\begin{equation}
 \widetilde\Phi^{(u)}(|\bm k|) =
 (1 - \alpha_{2(3q)}) \biggl[U^2(|\bm k|) + W^2(|\bm k|)\biggl]
 + \alpha_{2(3q)} {8\pi{\rm x}(1 - {\rm x}) \over E_{\bm k}}
 G_{2(3q)}({\rm x}, \bm k_\bot)~.
\label{Phi:6q}
\end{equation}
 The parameter $\alpha_{2(3q)}$ is the probability for a non-nucleon
 component in the deuteron which is a state of two colorless $(3q)$
 systems.
\begin{equation}
  G_{2(3q)}(x, \vec k_\bot) = {b^2 \over 2\pi} 
  {\Gamma(A + B + 2) \over \Gamma(A + 1)\Gamma(B + 1)}
  x^A (1 - x)^B ~ {\mbox{e}}^{-b k_\bot}.
\label{6q}
\end{equation}
\begin{wrapfigure}{l}{8cm}
\epsfig{figure=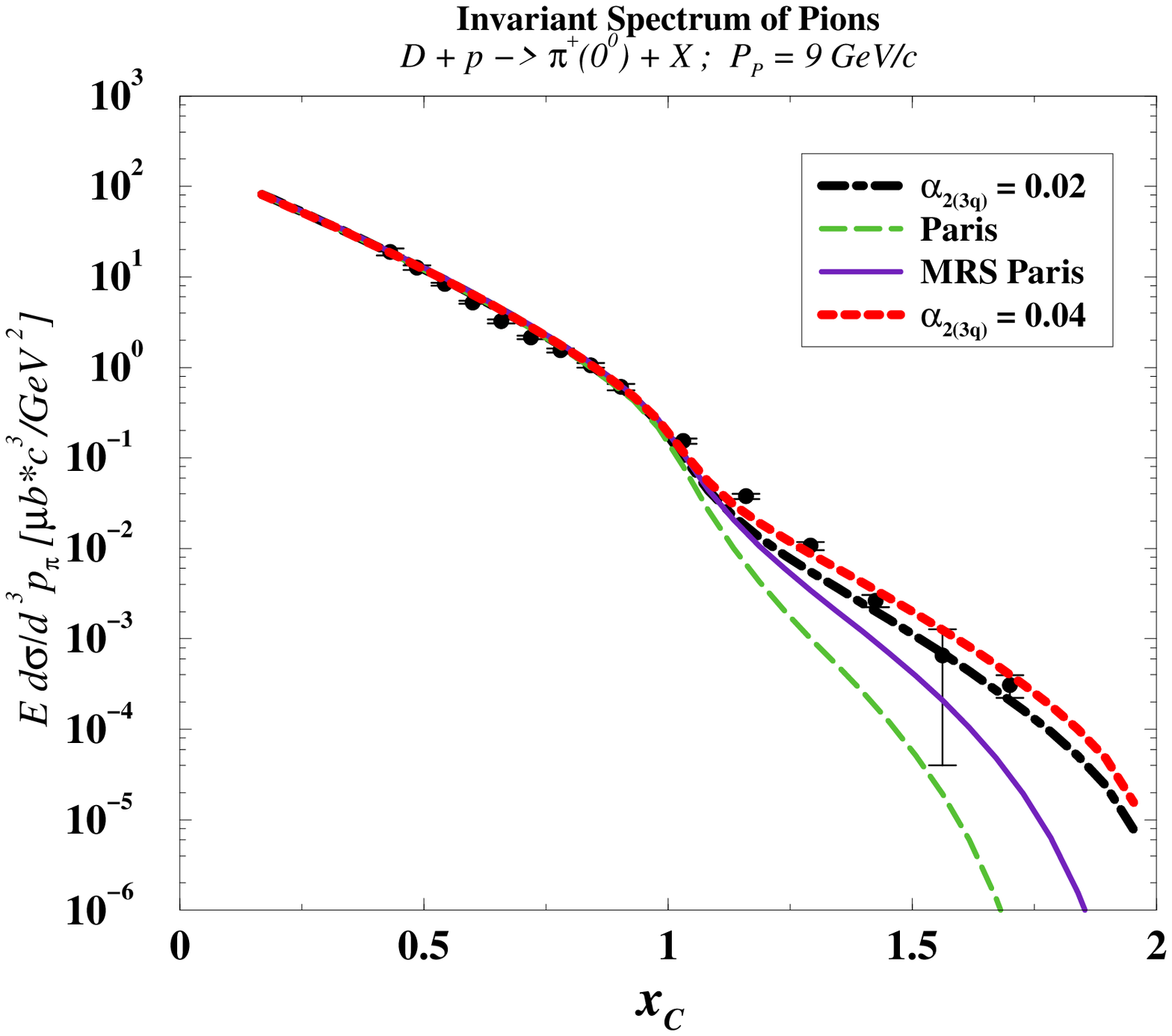, width=7.8cm, height=9cm}
{\small Figure 1:
 The invariant pion spectrum \protect\cite{bal85} calculated within the
 relativistic impulse approximation with including of the non-nucleon
 component in the DWF \protect\cite{efr88,lyk93}; its probability
 $\alpha_{2(3q)}$ is $0.02\div0.04$ (dot-dashed and dashed curves,
 respectively).
}
\medskip
\end{wrapfigure}
 Fig.~1 presents the invariant pion spectrum calculated within
 the relativistic impulse approximation including the non-nucleon component
 in the DWF \cite{lyk93,efr88}. One can see a good description of the
 experimental data \cite{bal85} at all cumulative variable $x_{\cal C}$.

 However in some sense, the information contained in $T_{20}$ from 
 $D p\rightarrow p X$ and $D p\rightarrow\pi X$ processes is redundant,
 because the main ingredient by analyzing these reactions within the
 impulse approximation are the same deuteron properties. Therefore   
 calculation of the tensor analyzing power including non-nucleon degrees
 of freedom in fragmentation of deuteron to pions can give us a new
 independent information concerning the deuteron structure at small
 $N-N$ distances. Actually, in \cite{efr88} a form of 
 $\widetilde\Phi^{(u)}(|\bm k|)$ has been constructed only. However,
 to calculate $T_{20}$ it is not enough, the corresponding orbital waves
 have to be known. Let us assume, the non-nucleon degrees of freedom
 result in a main contribution to the ${\cal S}$- and ${\cal D}$-waves of the
 DWF. Constructing new forms of these waves by including the
 non-nucleon degrees of freedom we have to require that the square of the
 new DWF has to be equal to the one determined by the Eq.(\ref{Phi:6q}).
 Introducing a mixing parameter $\alpha = (\pi/4)a$ one can find the forms 
 of new ${\cal S}$- and ${\cal D}$-waves as the following:
 \begin{eqnarray}
  \widetilde{U}(|\bm k|) &=&
   \sqrt{1-\alpha_{2(3q)}}U(|\bm k|) + \cos(\alpha) \Delta(|\bm k|)~;
 \label{def:Unew} \\
  \widetilde{W}(|\bm k|) &=&
   \sqrt{1-\alpha_{2(3q)}}W(|\bm k|) + \sin(\alpha) \Delta(|\bm k|)~,
 \label{def:Wnew}
 \end{eqnarray}
 where the function $\Delta$ has been obtained from the equation:
 \begin{equation}
  \widetilde\Phi^{(u)}(|\bm k|) ~=~
   \widetilde{U}^2(|\bm k|) + \widetilde{W}^2(|\bm k|).
 \label{rel:UW-Phi}
 \end{equation}

\begin{wrapfigure}{l}{8cm}
\epsfig{figure=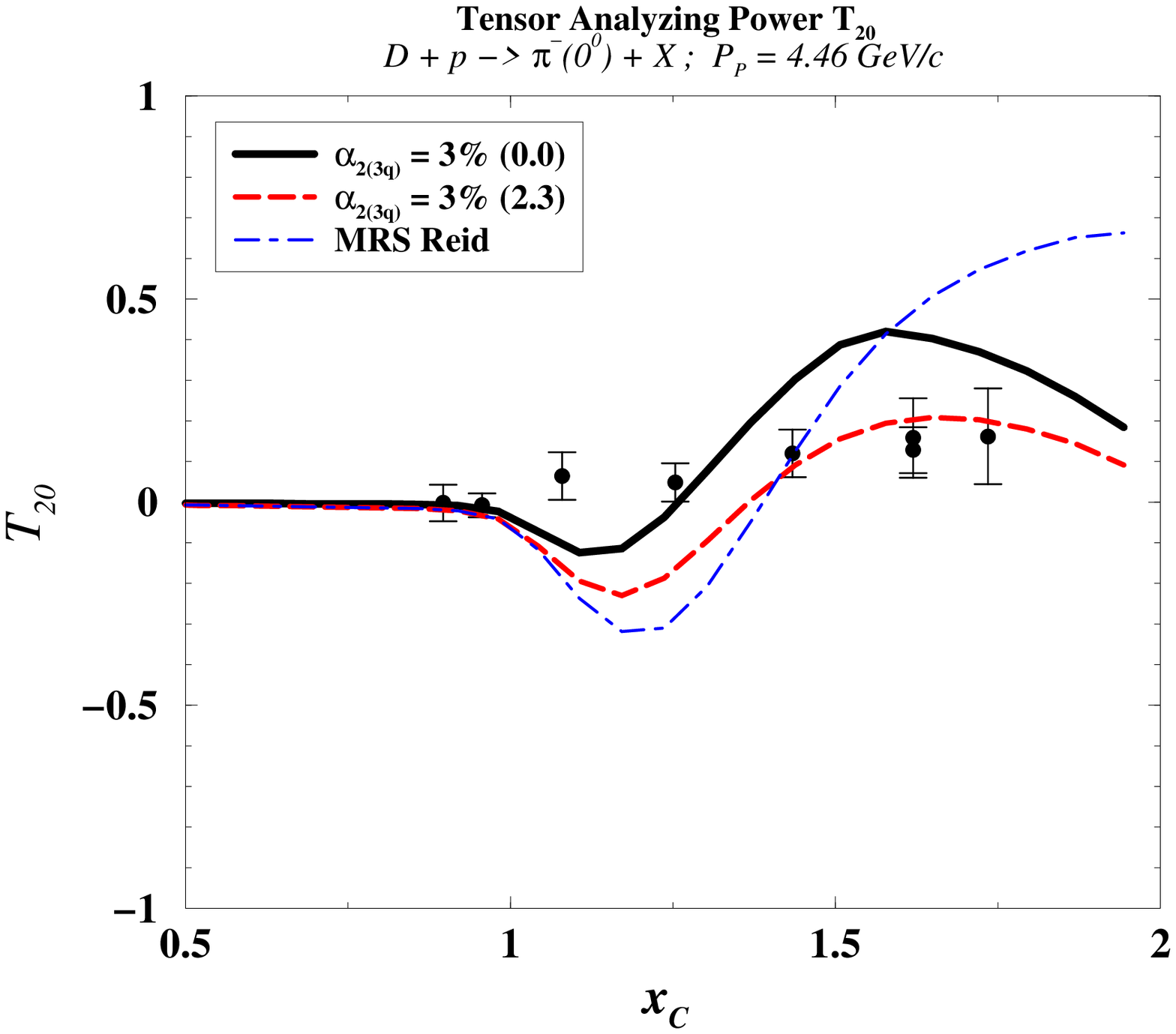, width=7.8cm, height=8cm}
{\small Figure 2:
 The tensor analyzing power $T_{20}$ \protect\cite{afa98} calculated within
 the relativistic impulse approximation including of the non-nucleon
 component in the DWF \protect\cite{efr88,lyk93} with probability
 $\alpha_{2(3q)} = 0.03$. The solid and dashed lines represent the
 calculations using the mixing parameter $a = 0.0$ and $2.3$, respectively,
 Eqs.~(\ref{def:Unew},\ref{def:Wnew}).
}
\medskip
\end{wrapfigure}
 Fig.~2 presents the analyzing power $T_{20}$ calculated by
 using the functions $\widetilde{U}, \widetilde{W}$ including the non-nucleon
 components in the DWF, according to \cite{lyk93,efr88}.
 It is shown from Fig.~2 the inclusion of non-nucleon components 
 in the DWF improves the description of the experimental data 
 about $T_{20}$ at $x_{\cal C} > 1$. The value of the parameter entering the
 Eqs.(\ref{def:Unew},\ref{def:Wnew}) $a = 2.3$ results in an optimal
 description of this observable.

 Main results can be summarized as the following. Very interesting
 experimental data on $T_{20}$ \cite{afa98} showing approximately 
 zero values at $x_{\cal C}\geq 1$  are not reproduced by a theoretical 
 calculus using even different kinds of the relativistic DWF \cite{ill00}. 
 This may indicate a possible existence of non-nucleon degrees of 
 freedom or basically new mechanism of pion production in the 
 kinematic region forbidden for free $N - N$ scattering.

 The inclusion of the non-nucleon degrees of freedom within the approach 
 suggested in \cite{efr88,lyk93} allows us to describe experimental data 
 about the inclusive pion spectrum at all the values of 
 $x_{\cal C}$ rather well, Fig.~1, and improve the description of data 
 \cite{afa98} concerning the analyzing power $T_{20}$ in the fragmentation 
 of deuteron to pions, Fig.~2.

\vspace{0.5cm}

The support from RFFI grant N99-02-17727 is gratefully acknowledged
by authors.

 \end{document}